\documentstyle[emulateapj,apjfonts,psfig]{article}

\lefthead{Miller, Fabian, \& Miller}  
\righthead{Cool Disks and IMBH Candidate ULXs}

\received{Date}
\revised{Date}
\accepted{Date}

\journalid{vol}{date}
\articleid{1}{4}
\paperid{id}

\cpright{AAS}{1999}
\ccc{x}

\begin{document}

\title{A Comparison of Intermediate Mass Black Hole Candidate ULXs and Stellar-Mass Black Holes}

\author{J.~M.~Miller\altaffilmark{1,2}, 
        A.~C.~Fabian\altaffilmark{3},
	M.~C.~Miller\altaffilmark{4}}

\altaffiltext{1}{Harvard-Smithsonian Center for Astrophysics, 60
	Garden Street, Cambridge, MA 02138, jmmiller@cfa.harvard.edu}
\altaffiltext{2}{NSF Astronomy and Astrophysics Fellow}
\altaffiltext{3}{Institute of Astronomy, University of Cambridge,
Madingley Road, Cambridge CB3 OHA, UK}
\altaffiltext{4}{Department of Astronomy, University of Maryland,
College Park, MD 20742}


\authoremail{jmmiller@cfa.harvard.edu}

\label{firstpage}

\begin{abstract}
Cool thermal emission components have recently been revealed in the
X-ray spectra of a small number of ultra-luminous X-ray (ULX) sources
with $L_{X} \geq 10^{40}$~erg/s in nearby galaxies.  These components
can be well fitted with accretion disk models, with temperatures
approximately 5--10 times lower than disk temperatures measured in
stellar-mass Galactic black holes when observed in their brightest
states.  Because disk temperature is expected to fall with increasing
black hole mass, and because the X-ray luminosity of these sources
exceeds the Eddington limit for $10~M_{\odot}$ black holes ($L_{Edd.}
\simeq 1.3\times 10^{39}$~erg/s), these sources are extremely
promising intermediate--mass black hole candidates (IMBHCs).  In this
Letter, we directly compare the inferred disk temperatures and
luminosities of these ULXs, with the disk temperatures and
luminosities of a number of Galactic black holes.  The sample of
stellar-mass black holes was selected to include different orbital
periods, companion types, inclinations, and column densities.  These
ULXs and stellar-mass black holes occupy distinct regions of a $L_{X}$
-- kT diagram, suggesting these ULXs may harbor IMBHs.  We briefly
discuss the important strengths and weaknesses of this interpretation.
\end{abstract}

\section{Introduction}
Ultraluminous X-ray sources (ULXs) are variable, off-nuclear, X-ray
point sources in nearby galaxies for which the implied luminosity of
the source exceeds the isotropic Eddington limit for a $10~M_{\odot}$
black hole (unless otherwise noted, in this work the term
``luminosity'' means the luminosity the source would have if it
radiates isotropically).  In the {\it Chandra} and {\it XMM-Newton}
era, a large number of these sources have been detected (see, e.g.,
Fabbiano \& White 2003, Miller \& Colbert 2004, Swartz et al.\ 2004).
The X-ray spectra and variability properties of the strong majority of
ULXs suggests that they are accreting sources.  These sources have
attracted a great deal of observational and theoretical attention, in
part because their luminosities suggest that they may harbor
intermediate-mass black holes (IMBHs; $M_{BH} \sim
10^{2-5}~M_{\odot}$).

As with any new class of sources, in the case of ULXs it was initially
tempting to identify the whole class as entirely one kind of source or
another.  The first {\it Chandra} study of the Antennae galaxies
suggested a population of IMBHs (Fabbiano, Zezas, \& Murray 2001), but
it was not long until a theoretical investigation suggested the ULXs
in the Antennae are stellar-mass sources (King et al.\ 2001); earlier
work suggested relativistic beaming in ULXs (Reynolds et al.\ 1997).  It may
be that ULXs, in particular those at the lower end of the ULX
luminosity distribution, are stellar-mass black holes (or even
neutron stars in rare cases).  However, a growing number of ULXs have
been identified which are strong IMBHCs.  Due in part to recent
observations which have obtained more sensitive spectra and
lightcurves, cool accretion disks have been found in some of the most
luminous ULXs ($L_{X} \geq 10^{40}$~erg/s).  Temperature is inversely
related to black hole mass ($T\propto M^{-1/4}$) in standard accretion
disks; the fact that these ULXs are 5--10 times brighter than
stellar-mass black hole candidates (BHCs) and yet have disks which are
5--10 times cooler than the disks in BHCs identifies them as
strong IMBHCs (see, e.g., Miller et al.\ 2003).

In this Letter, we directly compare the luminosity and disk
temperature of these IMBHC ULXs, with a number of well-known
stellar-mass BHCs.  This sample of ULXs indeed appears to be a distinct,
and perhaps rather homogeneous set of sources consistent with
harboring IMBH primaries.

\section{Data Selection}
\subsection{Intermediate Mass Black Hole Candidates}
We selected ULXs for which published fluxes imply luminosities of
$L_{X} \geq 10^{40}$~erg/s, and for which a soft thermal component is
required at the 3$\sigma$ level of confidence (or higher) in the
low-energy part of an X-ray spectrum which requires two continuum
components.  Six sources satisfy these selection criteria: NGC 1313
X-1, NGC 1313 X-2, M81 X-9 (Holmberg IX X-1), NGC 4559 X-7, Holmberg
II X-1, and NGC 4038/4039 X-37 (Antennae X-37).  See Table 1 for a
list of references for these sources.  M82 X-1 is the most luminous
ULX known, and in some respects it may be the single best IMBHC ULX
(see, e.g., Strohmayer \& Mushotzky 2003).  However, it lies in a
region of significant diffuse emission, and its low energy spectrum is
poorly determined.  Therefore, we have not included this source in our
comparison.

\subsection{Black Hole Candidates}
In an effort to prevent possible biases and to ensure a representative
sample, we selected stellar-mass black hole candidates which cover a
range of binary inclinations, primary masses, companion types,
distances, and absorbing columns.

\centerline{~\psfig{file=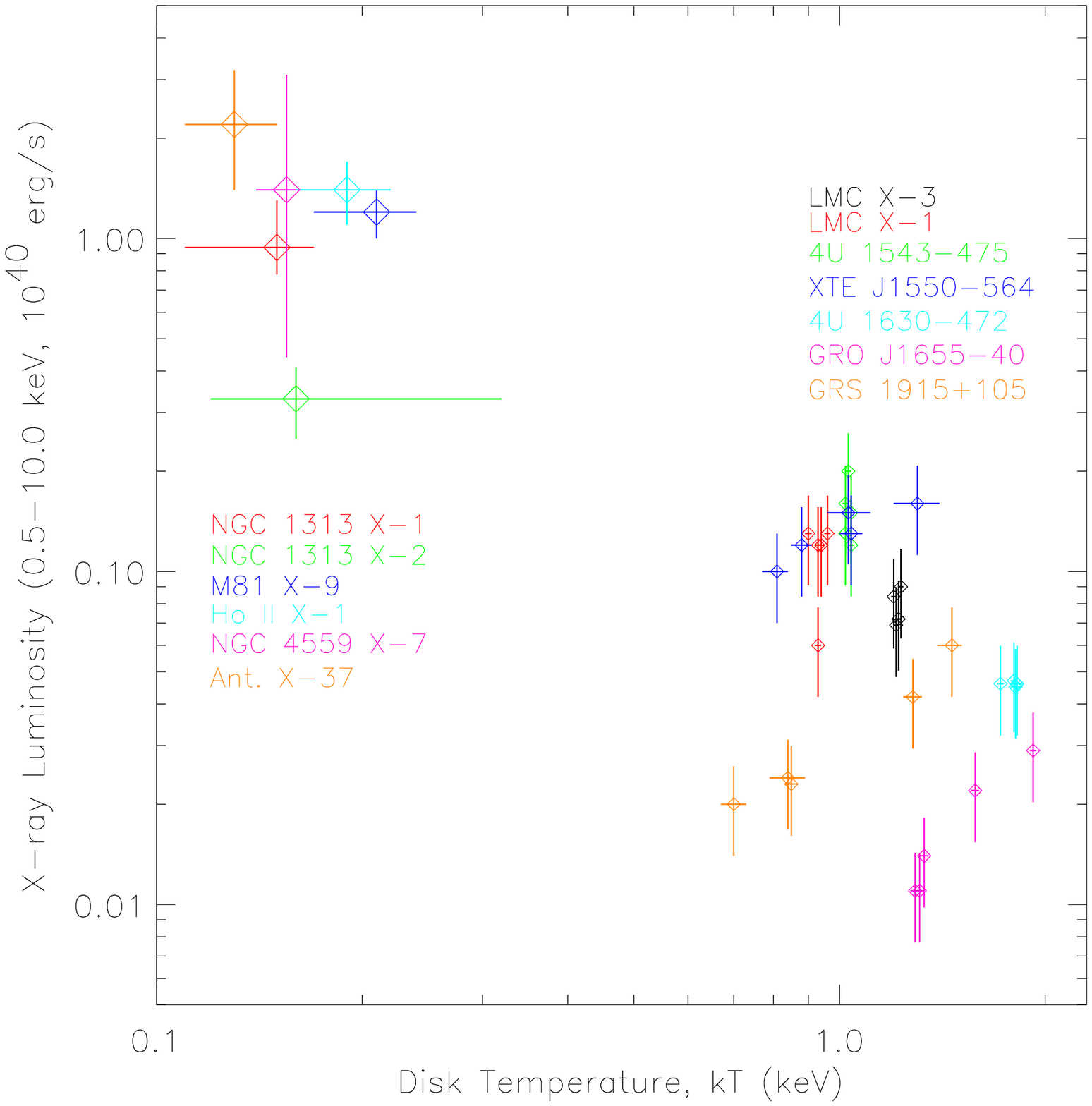,width=3.5in}~}
\figcaption[h]{\footnotesize In the figure above, the X-ray luminosity
of a number of extremely luminous ULXs and stellar-mass black holes in
their brightest phases are plotted against disk temperatures inferred
from X-ray spectral fits.  The fact that the ULXs are more luminous
and yet have cooler disks than the stellar-mass black holes suggests
that they may harbor intermediate mass black holes.}
\medskip

\noindent This sample includes: LMC X-1, LMC X-3, 4U 1543$-$475,
XTE~J1550$-$564, 4U 1630$-$472, GRO~J1655$-$40, and GRS~1915$+$105.
These are among the best-studied stellar-mass BHCs.  It should also be
noted that this sample includes both persistent sources or sources
undergoing very long outbursts (LMC X-1, LMC X-3, GRS 1915$+$105), and
transient sources (4U 1543$-$475, XTE J1550$-$564, 4U~1630$-$472, GRO
J1655$-$40) with outbursts which may last as much as a year (or
longer) followed by quiescent periods with fluxes 5--6 orders of
magnitude lower, lasting months to years.  See Table 1 for a list of
references for these sources.

\section{Analysis and Results}
The luminosity and disk temperatures we use in this paper are those
derived by the authors in the references listed in Table 1 from their
spectral fits to each source using a simple and phenomenological model
consisting of a multicolor disk blackbody (MCD; Mitsuda et al.\ 1984)
and power-law spectral components (both modified by neutral
interstellar absorption).  

The energy range over which the spectral fits were made to the ULX
sources and stellar-mass black hole sources differed considerably, due
to the different lower energy thresholds of {\it Chandra} and {\it
XMM-Newton}, and {\it RXTE}.  In most cases, the ULX spectra were fit
in the 0.2--10.0~keV or 0.3--10.0~keV range.  In contrast, the
stellar-mass black hole spectra were generally fit in the
3.0--25.0~keV or 3.0--100.0~keV range.  To perform a meaningful
comparison between these sources, we converted the flux and luminosity
measurements in the differing energy ranges to the 0.5--10.0~keV
range.  This conversion was accomplished by entering the exact
spectral model for each published spectral fit into XSPEC version 11.2
(Arnaud \& Dorman 2000), and measuring the ``unabsorbed'' flux of each
model in the 0.5--10.0~keV range.  For the ULXs in particular, the
0.2--100.0~keV luminosity --- more representative of a bolometric luminosity ---

\centerline{~\psfig{file=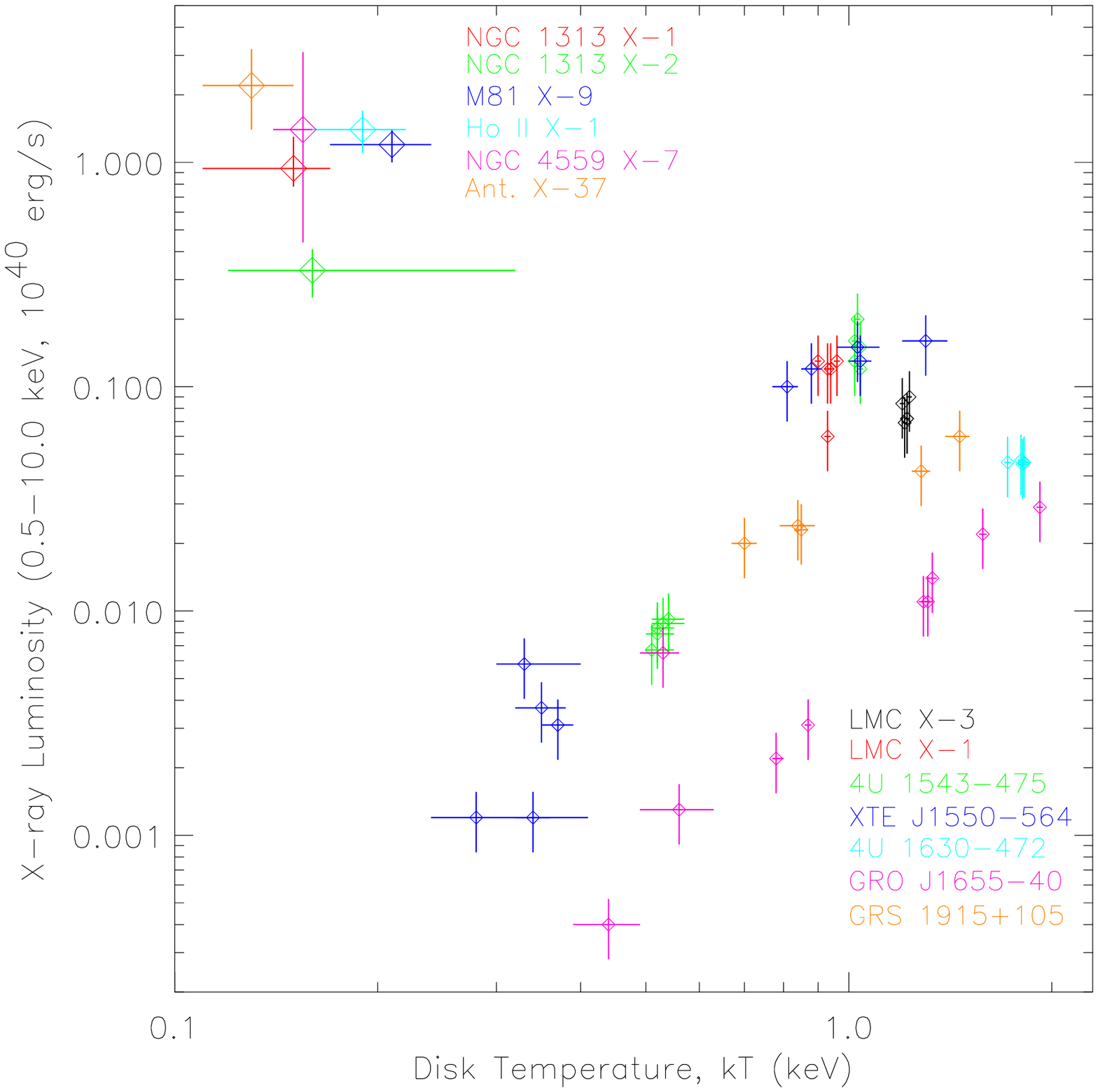,width=3.5in}~}
\figcaption[h]{\footnotesize In the plot above, a number of
low-luminosity BHC data points have been added to Figure 1, to
illustrate that BHC disk temperatures only approach those of the ULXs
at luminosities that are generally 2 orders of magnitude (or more)
below the ULX luminosities.}
\medskip

\noindent is a few times higher than the 0.5--10.0~keV
luminosity.  In some cases, the published disk temperatures were
``effective temperatures'' --- converted from ``color temperatures''
by application of a color correction factor, which attempts to account
for effects such as spectral hardening from radiative transfer through
a disk (Shimura \& Takahara 1995; Merloni, Fabian, \& Ross 2000;
Makishima et al.\ 2000).  In these cases, we converted the effective
temperature to a color temperature.  Color temperatures are compared
directly to color temperatures in this work.  This introduces no
significant temperature bias; a recent study has shown that the
correction factor for IMBHs should be very similar to that sometimes
applied to stellar-mass black holes (Fabian, Miller, \& Ross 2004).

The lower energy bound of the 0.5--10.0~keV range is somewhat
higher than the present lower energy bounds of {\it Chandra} and
{\it XMM-Newton}; however, history suggests that as all X-ray
detectors age, the lower energy bound gradually increases.  This range
was chosen to be forward--looking, and to avoid any biases inherent in
relying too heavily on the lowest bins in the {\it Chandra} and {\it
XMM-Newton} bandpasses.

To understand the properties of the ULX sources within the context of
BHCs in their brightest states, we plotted the luminosity and disk
temperature of each ULX, and the corresponding data for the five most
luminous observations of each BHC in our sample (see Figure 1).  In
selecting the brightest BHC observations, we are attempting to select
those phases wherein each source is closest to its Eddington
luminosity.  In all cases, the errors on the disk color temperatures
are 90\% confidence errors.  For the ULX sources, the luminosity
errors are the 90\% confidence errors in the measured flux.  A review
of the literature shows that constraints on the distance to given
Galactic sources can change considerably over time with refined
measurements, especially when extinction is particularly high.  To be
conservative, the luminosity errors on the stellar--mass black holes
were set by taking the best-fit measured flux, and enforcing
fractional errors of $\pm 30$\%.  This value is somewhat arbitrary,
but certainly greater than most quoted errors, and represents a
best-guess value based on the broad literature.

The difference between the IMBHC ULXs and standard, stellar-mass BHCs
is shown clearly in Figure 1.  The ULXs are generally 5--10 times more
luminous than the BHCs, and have inner disk color temperatures 5--10
times lower than the BHCs (it should be noted that while it is
possible to make a relatively cool disk appear hotter --- e.g., via
Compton-upscattering in some hot material --- it is not possible to make
an intrinsically hot disk appear cool).  Equally importantly, the ULXs
in Figure 1 are clustered together, suggesting they are fundamentally
similar.  In those BHC sources which span a reasonable range in
$L_{X}$, it is clear that $kT$ and $L_{X}$ (a trace of $\dot{m}$) are
positively correlated.  It is expected that the inner disk temperature
should be positively correlated with $\dot{m}$ (see, e.g., equation
5.43 in Frank, King, \& Raine 2002).  If a BHC is to reach the
luminosity window occupied by these ULXs, then, it is expected that
its disk temperatures should increase accordingly.  Indeed, XTE
J1550$-$564 was initially famous for flaring to 6.8~Crab --- a factor
of a few brighter than the highest points in Figure 1.  In an
observation which occurred within that flare, Sobczak et al.\ (2000)
measured an inner disk color temperature in excess of 3~keV.

It is interesting to explore the origin of the separation between the
ULXs and BHCs in $L_{X} - kT$ shown in Fig.\ 1.  To understand where
BHCs lie in this space when their disk temperatures approach those
seen in the ULXs, in Fig.\ 2 we have added the five data points
with the lowest disk temperatures from each BHC in the high luminosity
sample with such data.

Figure 2 clearly shows that when stellar-mass BHC inner disk
temperatures approach those seen in the ULX sample, their luminosity
has decayed to $few \times 10^{37}$~erg/s -- generally two orders of
magnitude (or more) below the luminosity of the ULXs.  Note that there
is a clear $L\propto T^{4}$ trend in the BHC data, as expected for
standard disks.  This plot also demonstrates that cool disk components
can be detected in Galactic stellar-mass black holes even with {\it
RXTE} (which has a low energy bound of 2~keV).  It is not the case
that very cool disks have not yet been detected in Galactic sources at
high luminosities because of instrumental thresholds, Galactic column
densities, or a combination of these.  At their highest luminosities,
stellar-mass black holes --- regardless of companion type, orbital
period, distance, or intervening absorption --- do not have disks as
cool as those found in IMBHC ULXs.

\section{Discussion}
The comparison undertaken in this work demonstrates that ULXs in our
sample are clearly different than our representative sample of
stellar-mass BHCs.  These ULXs are more luminous but have cooler
thermal disk components than standard stellar-mass BHCs; these facts
can be explained naturally if the ULXs harbor IMBHs.  The
comparison presented here makes the distinction more concrete, and
demonstrates that the differences are not due to instrumental effects
(e.g., detector energy thresholds), observational effects (e.g.,
column density), or astrophysical effects (e.g., companion
type or orbital period).

Optical, radio, and even X-ray data (see Pakull \& Mirioni 2003;
Miller et al.\ 2003; Miller, Fabian, \& Miller 2004; Strohmayer \&
Mushotzky 2003) suggest that the high luminosity of these ULXs cannot
easily be explained through funneling in the inner disk (e.g. King et
al.\ 2001), through relativistic beaming (e.g. Reynolds et al.\ 1997;
Kording, Falcke, \& Markoff 2002), or through an alternative
accretion flow geometries wherein a photosphere and shocks are
postulated instead of the conventional disk and corona (King
2003).  

This comparison further demonstrates the problems with present
theoretical alternatives to IMBH primaries in our ULX sample.  In
Figure 1 and Figure 2, it is clear that the stellar-mass BHC 4U
1543$-$472 ($M = 9.4\pm 2.0 M_{\odot}$, Orosz et al.\ 2004) exceeds
its {\it isotropic} Eddington limit; the source luminosity would be a
factor of a few higher still if the energy band considered extended
either down to 0.1 keV or up to 100.0~keV.  4U~1543$-$375 has an
inclination of $21^{\circ}$ (Orosz et al.\ 2004) and funneling might
be a means to introduce sufficient anisotropy to avoid violating the
Eddington limit.  It is also possible that a ``slim disk'' solution
may allow a luminosity in excess of the isotropic Eddington limit
(Watarai, Mizuno, \& Mineshige 2001; Begelman 2002).  It is important
to realize that although either explanation for 4U 1543$-$475 may
hold, 4U 1543$-$475 is not observed to have an anomalously low inner
disk color temperature.
  
The clustering of the ULXs in Figure 1 and Figure 2 suggests that they
may be fundamentally similar.  The similarity may be that these
sources harbor IMBHs with masses that lie in a rather narrow range.
It is difficult to identify the mass range implied for these sources
precisely: scaling the ULX luminosities to the isotropic Eddington
luminosity for a $10~M_{\odot}$ black hole is sensitive to the energy
range on which the luminosities are inferred, and scaling the ULX
inner disk color temperatures to those seen in stellar-mass BHCs is
sensitive to the temperature assumed to be typical for those sources
when they accrete near to their Eddington limit.  For these ULX
sources, a reasonable mass range may be $100-3000~M_{\odot}$ (see,
e.g., Miller et al.\ 2003; Miller, Fabian, \& Miller 2004; Cropper et
al.\ 2004).  If they are accreting at approximately one tenth of their
Eddington limits, lower masses implied by Eddington limit scaling
would come more into line with the high mass estimates that come from
scaling the ULX inner disk color temperatures to an inner disk color
temperature of 1~keV in BHCs.

The comparisons in this analysis present a strong case for IMBHs in a
few ULXs, but it is worth addressing some ways in which this
interpretation may ultimately be proved incorrect.

The soft thermal component in these ULX spectra have been fit with
disk models.  This is because there are no compelling soft X-ray
emission lines (individual lines significant at the 3$\sigma$ level or
higher, excluding Fe~K$\alpha$ lines which are likely due to disk
reflection) yet reported in any ULX spectrum.  Thus, thermal plasma
models are not statistically required, and simpler spectral forms are
assumed.  However, even in spectra with moderate sensitivity (the ULX
spectra so-far obtained are certainly of moderate sensitivity), it is
difficult to statistically rule-out thermal plasma models.  Matters are
further complicated by the fact that some ULX lie near to star-forming
regions, where a thermal plasma may be present but unrelated to the
source (M82 X-1 is a good example; see Strohmayer \& Mushotzky 2003).
Although the contribution of a thermal plasma to the soft excess in
these ULXs would seem to be small, improved spectra are needed to
tightly constrain any such contribution.

Soft excesses have been found in a number of AGN, which have
temperatures similar to those seen in the ULXs in this sample when fit
with a disk model (much too hot for such massive black holes; see
Gierlinski \& Done 2004).  The origin of the soft excess in these AGN
--- and whether or not it is due to a disk --- remains uncertain.  The
``big blue bump'' would seem to be far more likely to be the primary
disk contribution in these AGN.  Photosphere plus shocks models are as
implausible in these sources as they are in ULXs (see Miller, Fabian,
\& Miller 2004).  It has also been suggested that the excess may be
due to relativistically--skewed soft X-ray absorption edges
(Gierlinski \& Done 2004) or relativistically-skewed disk reflection
features (Fabian et al.\ 2004).  Although it has been suggested that
ULXs in elliptical galaxies may only be background AGN (Irwin,
Bregman, \& Athey 2004), the proximity of the ULXs in this sample to
their galactic nuclei, star-forming regions, or spiral arms suggests
that they are properly associated with their host galaxies and not
background sources.  Even though these IMBHC ULXs are unlikely to be
background AGN with soft components, the difficulties found in
understanding the soft X-ray excesses in a number of AGN illustrates
that the spectral continuum is not well-understood in all accreting
sources. \\

J. M. M. acknowledges support from the NSF through its Astronomy and
Astrophysics Postdoctoral Fellowship program.  M. C. M. was supported
in part by NSF grant AST 00-98436 and NASA grant NAG 5-13229.


\begin{table}[t]
\caption{ULX and BHC Information and References}
\begin{footnotesize}
\begin{center}
\begin{tabular}{lllllllll}
Source Name & T/P$^{a}$ & Companion & P (hr) & distance & $N_{H}$ & $\theta_{i}$ & High $L_{X}$$^{b}$ & Low $L_{X}$$^{c}$\\
~ & ~ &~ & ~ & ~ & ($10^{21}~{\rm cm}^{-2}$) & ~ &~ & ~ \\

\tableline

NGC 1313 X-1 & -- & -- & -- & 3.7~Mpc$^{1}$ & 4.4$^{1}$ & -- & -- & -- \\

NGC 1313 X-2 & -- & -- & -- &3.7~Mpc$^{1}$ & 3.0$^{1}$ & -- & -- & -- \\

M81 X-9 & -- & -- & -- & 3.4~Mpc$^{2}$ & 2.8$^{2}$ & -- & -- & -- \\

Ho II X-1 & -- & -- & -- & 3.4~Mpc$^{3}$ & 1.6$^{3}$ & -- & -- & -- \\

NGC 4559 X-7 & -- & -- & -- & 9.7~Mpc$^{4}$ & 5.1$^{4}$ & -- & -- & -- \\

Antennae X-37 & -- & -- & -- & 19~Mpc$^{5}$ & 5.6$^{5}$ & -- & -- & -- \\

\tableline

LMC X-1 & P & O7III$^{6}$ & 101.5$^{6}$ & 50~kpc$^{7}$ & 7.2$^{7}$ & -- & (3,21,25,27,30)$^{7}$ & -- \\

LMC X-3 & P & B3V$^{6}$ & 40.9$^{6}$ & 50~kpc$^{7}$ & 0.32$^{7}$ & -- & (5,6,7,8,9)$^{7}$ & -- \\
 
4U 1543$-$475 & T & A2V$^{6}$ & 26.8$^{6}$ & 7.5~kpc$^{6}$ & 4.0$^{8}$ & 21$^{\circ}$$^{9}$ & (4,5,6,7,8)$^{8}$ & (45,46,47,48,49)$^{8}$ \\

XTE J1550$-$564 & T & G8IV--K4III$^{6}$ & 37.2$^{6}$ & 5.3~kpc$^{6}$ & 20$^{10}$ & 72$^{\circ}$$^{9}$ & (17,18,19,20,21)$^{10}$ & (193,195,196,200,206)$^{10}$ \\

4U 1630$-$472 & T & -- & -- & 8.5~kpc & 90$^{11}$ & -- & (28,29,30,32,34)$^{11}$ & -- \\

GRO J1655$-$40 & T & F6III$^{6}$ & 62.4$^{6}$ & 3.2~kpc$^{6}$ & 8.9$^{12}$ & 70$^{\circ}$$^{9}$ & (725,801,806,815,816)$^{12,13}$ & (729,803,814,818,825)$^{12,13}$ \\
 
GRS 1915$+$105 & P & K--MIII & 804.0 & 11~kpc$^{15}$ & 4.0--4.8$^{14}$ & 66$^{\circ}$$^{15}$ & (1,2,3,4,5)$^{14}$ & -- \\

\tableline
\end{tabular}
\end{center}
\tablecomments{
$^{a}$ Denotes whether the source is transient or persistent.
$^{b}$ For BHCs, denotes which high luminosity observations were selected.
$^{c}$ For BHCs, denotes which low luminosity observations were selected.
Where a value is not given, it is unknown for the given source.
$^{1}$ Miller et al.\ (2003).
$^{2}$ Miller, Fabian, \& Miller (2004).
$^{3}$ Dewangan et al.\ (2004).
$^{4}$ Cropper et al.\ (2004).
$^{5}$ Miller et al.\ (2004).
$^{6}$ McClintock \& Remillard (2004).
$^{7}$ Wilms et al.\ (2001).
$^{8}$ Park et al.\ (2004).
$^{9}$ Garcia et al.\ (2003).
$^{10}$ Sobczak et al.\ (2000).
$^{11}$ Trudolyubov et al.\ (2001).
$^{12}$ Sobczak et al\ (1999).
$^{13}$ Date in MJD, minus 960,000.
$^{14}$ Ueda et al.\ (2002).
$^{15}$ Fender et al.\ (1999).
}
\vspace{-1.0\baselineskip}
\end{footnotesize}
\end{table}

\end{document}